# Magnetic Vortex Core Dynamics in a Ferromagnetic Dot

K. Yu. Guslienko,[1*] X.F. Han,[2] D.J. Keavney,[2] R. Divan,[2] and S. D. Bader[1]

[1] *Materials Science Division, Argonne National Laboratory, Argonne, IL 60439*
[2] *Advanced Photon Source, Argonne National Laboratory, Argonne, IL 60439*

Abstract

We report direct imaging by means of x-ray photoemission electron microscopy of the dynamics of magnetic vortices confined in micron-size circular Permalloy dots that are 30 nm thick. The vortex core positions oscillate on a 10-ns timescale in a self-induced magnetostatic potential well after the in-plane magnetic field is turned off. The measured oscillation frequencies as a function of the aspect ratio (thickness/radius) of the dots are in agreement with theoretical calculations presented for the same geometry.

*[)] Author to whom correspondence should be addressed. Electronic address:gusliyenko@anl.gov;

Ultrasmall, ultrafast phenomena are at the forefront of condensed matter research presently, and magnetism provides opportunities to explore this realm. A new era in the investigation of mesoscopic magnetic structures started recently due to advances in nanolithography, which offer a unique opportunity to prepare a variety of novel laterally-confined nanostructured magnets. Patterned nanomagnets (see Ref. 1 and references therein) are promising candidates for ultrahigh density discrete recording media, magnetic random access memory cells, miniature magnetic field sensors and spintronic logic devices. The lowest total energy configuration of lithographically patterned micron- and sub-micron soft magnetic particles in remanence can be a vortex state [2, 3]. Vortices in magnetic systems have been observed by magnetic force [4], Lorentz [5], spin-polarized scanning tunneling [6] and X-ray microscopies [7-9] and manipulated by a magnetic field [10]. The vortex consists of a core with magnetization perpendicular to the dot plane and the main part with an in-plane flux-closure magnetization distribution [4-6, 10]. The vortex ground state leads to a complex spin excitation spectrum, with a low-frequency mode of displacement of the vortex as a whole [11]. The classification of eigenmodes of a vortex and the influence of the magnetic dipolar interaction on vortex excitations in particles of small size has attracted much attention recently [12-15]. The vortex translation, radial and azimuthal modes have been explored theoretically for small cylindrical dots. In general, their spin excitation spectrum is dominated by strong magnetostatic contributions defined by the dot geometry. Magnetic vortex excitations have a special significance because they control the magnetization reversal of novel spintronic devices.

Vortex excitations observed by Brillouin light scattering (BLS) [12, 13] have relatively high frequencies and can be interpreted as quantized spin waves in restricted geometries [13-15]. BLS, however, does not permit the detection of eigenfrequencies in the sub-GHz range. The frequency of the vortex motion is generally too low to be detected by means of standard ferromagnetic resonance cavity techniques. The lowest frequency translational (or gyrotropic) mode can be excited by an in-plane pulsed magnetic field [16] and was observed by time-resolved magneto-optical Kerr measurements [17] and by microwave absorption techniques [18]. It has an eigenfrequency ~ 100 MHz for dots with thicknesses much less than their diameter. Remarkably, the finite value of the eigenfrequency is a result of the finite dot size and is determined not by exchange but by long range dipole-dipole (magnetostatic) interactions. Photoemission electron microcopy (PEEM) techniques that utilize x-ray magnetic circular dichroism (XMCD) to obtain the magnetic contrast is emerging as one of the premier probes [7, 8]. The vortex core trajectory induced by pulsed magnetic fields has recently been detected in square dots by PEEM imaging. Raabe *et al*. [9] conducted XMCD/PEEM dynamics experiments in squares and did not observe the gyrotropic mode. Moreover, they observed that the vortex core displacement is perpendicular to the exciting magnetic field, which is in contradiction with the experiment by Choe *et al*. [7] who observed initial core shift parallel to the pulse field followed gyrotropic motion. We intend to clarify these discrepancies. The experiments in Ref. [7, 8] did not track the eigenfrequencies of the vortex core oscillations and their geometric dependences, but instead detected parts of the vortex core trajectories [7] or provided vortex images [8]. There are domains and domain walls in square elements that reveal their own response to pulse fields [9] and that can mask the oscillations of the vortex core. The data provide qualitative information about the low frequency region of the vortex excitation spectrum. Hence, we chose to probe circular dot structures in order to avoid these complications and observe core trajectories and eigenfrequencies in a transparent form.

The present work combines PEEM and theoretical modeling to demonstrate the existence of the vortex translational mode in cylinder-shaped dots and improve our fundamental understanding of the magnetization dynamics of vortices trapped in small ferromagnetic particles. We report XMCD/PEEM detection of the low-frequency vortex excitations for micron-sized $Fe_{20}Ni_{80}$ dots as a function of the geometrical aspect ratio. The spatial resolution is ~300 nm and time resolution ~ 0.1 ns, which permits the detection of vortex position oscillations with high accuracy. A moving magnetic vortex experiences a force (gyroforce) perpendicular to its velocity [19] and a magnetostatic force due to the restricted geometry. We observed free vortex oscillations after field pulse for times about 100 ns that was enough to detect the vortex core oscillation periods, which were estimated for our samples to be ~ 20 ns.

The experiments were performed at the soft X-ray beamline (4-ID-C) in sector 4 of the Advanced Photon Source (APS). This beamline provides left and right circularly polarized x-rays with 96% polarization in the energy range 500 - 3000 eV. Images were obtained with an Omicron-Focus PEEM to record the absorption coefficient across the sample surface at the Ni $L_3$ resonance (852.7 eV). By taking the difference of the images acquired with left- and right-circularly polarized light, topographical and chemical contrast can be largely eliminated to yield only the magnetic contrast. The intensity in the difference images is proportional to the dot product $(\mathbf{M} \cdot \mathbf{k})$ of the magnetization vector $\mathbf{M}$ and x-ray propagation direction $\mathbf{k}$. The experimental setup is schematically shown in Fig. 1. X-rays are incident on the samples at 25° with respect to the surface. Using the coordinates defined in Fig. 1, the projection of the x-ray propagation direction onto the sample surface plane is along the $x$ direction. Therefore, we are mainly sensitive to the magnetization component along the $x$ direction. We used the standard 24-bunch fill pattern of the APS storage ring, in which the temporal spacing of the x-ray pulses is 153 ns. The standard deviation of the bunch length is 40-ps [20], which limits the best possible temporal resolution to ~86 ps. Triggered by the timing signal provided by the storage ring with an adjustable time delay, a commercial pulse generator launches current pulses to a coplanar waveguide (CPW) fabricated via photolithography, sputtering and liftoff techniques. The CPW consists of a 1-µm thick gold on a Si wafer with nominal 50 Ω characteristic impedance. On top of the center track of the CPW, a 30-nm thick Permalloy film capped with a 2-nm thick Cr is patterned via e-beam lithography and liftoff into circular shapes with diameters of 6.3, 5.3 and 4.3 µm. This configuration favors excitation of the translational mode of the magnetic vortex. The films were sputtered in the absence of a magnetic field in a chamber of 8 $10^{-8}$ Torr base pressure. The waveform of the current pulses was measured at the output of the CPW with a 20-GHz sampling scope. The fall time of the current pulses is 240 ps. The magnetic field applied to the sample along the $x$ direction ($H_x$), created by the current pulses, can be estimated to be $I/2w$, where $I$ is the current and w=10 μm is the width of the center track. The estimated magnitude of $H_x \sim 30$ Oe is shown in the lower right of Fig. 1.

PEEM images for the 6.3 μm diameter dot are presented in upper part of Fig.2. Figures 2(A), 3 show the time dependence of the Y(t) displacement of the vortex core for the dots with different diameters. The X, Y displacements were extracted from images in the following straightforward way based on the fact that the intensity in XMCD/PEEM images is proportional to the component $M_x = (\mathbf{M} \cdot \mathbf{k})/k$. The intensity along every horizontal line of the images was averaged. Where the average intensity changes sign should be where the vortex core is located. From Figs. 2-3, we make the following two observations. First, after the external field was turned off around 0 ns, the $Y$ displacement of all the three dots exhibits well defined oscillations with magnitude of 400-600 nm. Secondly, the oscillation frequency increases as the dot diameter decreases. The data in Fig. 2(A) and the selected images on the top of Fig. 2 are related by letters (a)-(l). From the images, one can see that in (a) the vortex core is above the center of the dot, in (b) it has moved to the center, in (c) it is below the center, and in (d) it has moved back to the center, and so on. Quantitative analysis of the data in Figs. 2, 3 yields the frequencies of these oscillations: 45.0 MHz (the diameter 2R=6.3 μm), 56.1 MHz (2R=5.3 μm) for a pulse length of 19 ns, and 43.4 MHz (2R=6.3 μm), 53.2 MHz (2R=5.3 μm), 63.3 MHz (2R=4.3 μm) for a 75-ns pulse length.

Magnetic vortices are a form of topological solitons [21, 22], *i.e.* topological excitations of an ordered system defined as a non-uniform configuration of the order parameter (magnetization) that cannot be transformed into the uniform configuration by any continuous deformation. Magnetic vortices are metastable states for 2D isotropic ferromagnets [23] having energy proportional to their vorticity. After pioneering work by Thiele [19] there were considerable efforts to develop theoretical models of magnetic vortices in infinite 2D ferromagnets taking into account the anisotropic exchange interaction (see Refs. 24 and 25 and references therein). To describe the vortex dynamics in the finite magnetic particles we consider a continuum model with 3D classical magnetic moments. A 2D magnetization distribution is assumed (independent of the *z*-coordinate along the dot thickness): $\mathbf{m}(\boldsymbol{\rho},t) = \mathbf{M}(\boldsymbol{\rho},t)/M_s$, $M_s$ is the saturation magnetization, $\boldsymbol{\rho} = (x,y)$. The description of the vortex oscillations is based on the effective equation for vortex collective coordinates [19, 24] derived from the Landau-Lifshitz equation of motion. The properties of magnetic vortices can be described by three integers: the topological charge *q* or vorticity, the polarization *p* of the core, and the chirality $C = \pm 1$ (counter-clockwise or clockwise rotation of the static vector **m** in the dot plane). Vorticity is degree of mapping of *x-y* plane to the surface of unit sphere $\mathbf{m}^2(x,y)=1$ determined by the function $\mathbf{M}(x,y)$ and the core polarization describes "up" or "down" direction of the normal magnetization component of the vortex core.

We consider here only the low-frequency vortex translation modes, where the vortex center oscillates about its equilibrium position. The magnetization distribution of a moving vortex is distinctively different from its static counterpart. The dynamic magnetization should also satisfy strong pinning boundary conditions $(\mathbf{m} \cdot \mathbf{n})_S = 0$ [26] on the side dot surface *S* (**n** is the normal vector to the side surface). We use the "surface-charge-free" spin distribution model to calculate the translational mode frequency. In this case, the magnetization configuration of an oscillating vortex is such that the magnetic "charges" on the particle's lateral surfaces are eliminated and the boundary conditions are satisfied. The model does indeed provide a realistic dynamic spin distribution because it corresponds to a minimization of the exchange energy and the main (side and face surfaces) contribution to the magnetostatic energy.

We consider a disk-shaped particle with thickness *L* and radius *R*. We use the angular parameterization for the dot magnetization components $m_z = \cos\Theta$, $m_x + im_y = \sin\Theta \exp(i\Phi)$ and define a topological invariant *q* (vorticity) for the given magnetization distribution [22] as $q = \int \sin\Theta(\rho)d\Theta(\rho)d\Phi(\rho)/2\pi$, where $\Theta(\rho)$, $\Phi(\rho)$ are the solutions of the Landau-Lifshitz equation. We use the Landau-Lifshitz equation of motion with the energy density $w = A[(\nabla\Theta)^2 + \sin^2\Theta(\nabla\Phi)^2] + w_m$, where $w_m = -\mathbf{M}\cdot\mathbf{H}_m/2$ is the magnetostatic energy density, $\mathbf{H}_m$ is the magnetostatic field, and $A$ is the exchange stiffness. The vortex center behaves like a particle and can be characterized by its coordinate $\mathbf{X}$, mass, momentum, *etc*. The parameters of the equations of motion for $\mathbf{X}$ depend on the vortex magnetization distribution. The generalized Thiele's equation of motion [19] for the vortex, accounting the vortex mass is:

$$\hat{M}\ddot{\mathbf{X}} - \mathbf{G}\times\dot{\mathbf{X}} + \partial_\mathbf{X}W(\mathbf{X}) = 0, \quad (1)$$

where $\mathbf{X}=(X, Y)$ is the vortex center position, $W(\mathbf{X})$ is the energy of the vortex shifted from its equilibrium position in the dot center $\mathbf{X}=0$. $\hat{M}$ is the vortex mass that is a tensor, in general. The dot over the symbol means the derivative with respect to time.

The first term in Eq. (1) is analogous to the classical acceleration term, but the vortex mass is small for typical micron size dots and hence can be safely neglected. The second term is the gyroforce $\mathbf{F}_g = \mathbf{G}\times\dot{\mathbf{X}}$ determined by the vortex non-uniform magnetization distribution [19]. The gyrovector $\mathbf{G} = -G\hat{\mathbf{z}}$ can be represented via components of the antisymmetric gyrotensor $\hat{G} = \int dV\hat{g}$ (non-zero components $G_{xy} = -G_{yx} = G$, $\hat{\mathbf{z}}$ is the unit vector perpendicular to the dot plane $xOy$), where the gyrocoupling density tensor $\hat{g}$ is $g_{\mu\nu} = (M_s/\gamma)\mathbf{m}\cdot[\partial_\mu\mathbf{m}\times\partial_\nu\mathbf{m}]$. The gyroconstant $G$ for a magnetic vortex is $G = 2\pi qpLM_s/\gamma$, where $\gamma$ is the gyromagnetic ratio. The non-zero gyrovector, an intrinsic property of the vortex arising from non-zero topological charges of the vortex core, is principally important for the vortex dynamics description. The vortex core dynamics is non-Newtonian due to the dominating contribution of the gyrotropic term in the equation of motion. The gyrovector can be calculated by integration over the vortex core, where $\nabla\Theta \neq 0$ and determines the direction of the core rotation and its frequency. The gyroforce in Eq. (1) is analogous to the Lorentz force acting on a moving electrical charge. The role of the charge is played by the vorticity $q$ of the vortex magnetization distribution.

The dynamic restoring force [third term in Eq. (2)] appears due to finite dot sizes and is directed toward the vortex equilibrium position. For micron and submicron dot radii the small exchange contribution $\sim AL$ can be neglected and the dot self-induced magnetostatic energy gives the main contribution to the dependence of the vortices energy $W(\mathbf{X})$ on the coordinates. To describe the vortex motion we need to introduce a description of the vortex shifted from the center of the dot. For this purpose we use the complex function [11] $w(\zeta,\bar{\zeta}) = \tan(\Theta(x,y)/2)\exp(i\Phi(x,y))$ of the complex variable $\zeta = (x+iy)/R$. Accounting for the constraint that $\mathbf{m}^2 = 1$, the function $w(\zeta,\bar{\zeta})$ can be presented as $w(\zeta,\bar{\zeta}) = f(\zeta)$ if $|f(\zeta)|<1$ (within the vortex core) and $w(\zeta,\bar{\zeta}) = Argf(\zeta)$ if $|f(\zeta)| \geq 1$, where $f(\zeta)$ is an analytical function. In the case of magnetic vortex motion in a cylindrical dot, the function is $f(\zeta) = (i/c)C(\zeta - s - \bar{s}\zeta^2)$, where $c=b/R$ is the normalized core radius, $s = (X+iY)/R$ is the vortex core position. The function $f(\zeta)$ describes the magnetization distribution as a superposition of two vortices, one centered within the dot at $z=s$ and the other is an image vortex located outside the dot at $s/|s|^2$. The function $f(\zeta)$ relates magnetization oscillations $\mathbf{m}(\mathbf{r},t)$ to the vortex center position oscillations $\mathbf{X}(t)$. The core occupies a relatively small area of the dot ($b\sim$ 10-20 nm [6]), therefore the vortex motion is determined mainly by magnetostatic forces due to dynamic magnetic charges outside the core. The vortex translation frequency calculation then is reduced to the calculation of the energy W($\mathbf{X}$). The main contribution to $W(\mathbf{X})$ is the magnetostatic energy of volume charges with density $\sigma = M_s \partial_\mu m^\mu(\mathbf{r},\mathbf{X})$:

$$W_m(\mathbf{X}) = \frac{1}{2} \int d^3\mathbf{r}\, d^3\mathbf{r}' \frac{\sigma(\mathbf{r},\mathbf{X})\sigma(\mathbf{r}',\mathbf{X})}{|\mathbf{r}-\mathbf{r}'|}, \qquad \sigma(\mathbf{r},\mathbf{X}) = \frac{2}{R} M_s \operatorname{Re}\left(\frac{\partial w(\zeta,\bar{\zeta})}{\partial \zeta}\right), \qquad (2)$$

where integration is conducted over the dot volume $V$, $\mathbf{m}(\mathbf{r},\mathbf{X})$ is the magnetization distribution of a shifted vortex centered at $\mathbf{X}$, and $\mathbf{r}=(\boldsymbol{\rho},z)$ is the radius vector.

For small ($|s| \ll 1$) displacement of the vortex center from its equilibrium position ($\mathbf{X}=0$) one can write the decomposition $W(\mathbf{X}) = W(0) + \kappa |\mathbf{X}|^2/2$, where the stiffness coefficient $\kappa$ can be determined from the energy of self-induced charges (2). The vortex core oscillates in a potential well (with minimum in the dot center $\mathbf{X}=0$) created by the magnetostatic energy with eigenfrequency $\omega_0 = G^{-1}\kappa$ and the core trajectory being a circle. The vortex chirality and polarization do not contribute either to the eigenfrequency or to the trajectory shape. The eigenfrequency of the vortex translational mode is given by:

$$\omega_0 = 8\pi\gamma M_s F(\beta), \qquad (3)$$

where $F(\beta) = \int_0^\infty dt\, t^{-1} f(\beta t) I^2(t)$ is a function of the dot aspect ratio $\beta = L/R$, $I(t) = \int_0^1 dx\, x J_1(tx)$, $f(x) = 1 - (1-\exp(-x))/x$.

The vortex translation frequency (3) in micron-size dots is determined mainly by magnetostatic interactions and depends only on the dot aspect ratio $\beta$. For our dots $\beta$ is within the range 0.009-0.014. The function $F(\beta) \approx (5/18\pi)\beta$ at $\beta \ll 1$ leads to a simple expression for the vortex eigenfrequency $\omega_0 = (20/9)\gamma M_s \beta$ for thin cylindrical dots. The calculated frequency (3) is plotted together with the experimental frequency of the core oscillations in Fig. 4 assuming $\gamma/2\pi$=2.95 MHz/Oe. $M_s$=720 G was calculated from the best fit that is typical value for Permalloy. The experimental frequencies are in quantitative agreement with the frequencies calculated from Eq. (3).

Although the theory predicts that the vortex core should undergo gyrotropic motion after the external field is turned off, from observing our PEEM images as a function of time, it is impossible to conclude if there are oscillations of the vortex core along the *x* direction. The experimental error is bigger for the *X* component and the oscillation period cannot be reliably detected. We can observe that the *X*-component is non-zero, large (few hundred nm, see Fig. 2(B)) and is comparable in magnitude to the *Y*-component. The major difference of our sensitivity to the *Y* and *X* displacement of vortex core comes from the fact that we are only detecting $M_x$. For instance, if one horizontal and one vertical line are drawn across the vortex core, along the vertical line, the intensity changes sign while along the horizontal line it is almost featureless. The presented model yields $m_x(x,y) = CR[-y + x(s_y x - s_x y)(R^2 - \rho^2)/R\rho^2]/\rho$. The line of zero-contrast $m_x(x,y) = 0$ is $y = s_y(1 - x^2/R^2)$ and does not depend on $s_x$ in the main approximation. Therefore, it is more difficult to detect an *X* core displacement and we extracted the vortex core eigenfrequency from *Y*(*t*) oscillations. *I.e.*, the observation method assumes considerably high sensitivity to the vortex core shift component perpendicular to the pulse field **H**∥**k**. This explains why the shift component parallel to the pulse field was not observed in Ref. 9. Choe et al. [7] explored field driven vortex motion with a frequency of 125 MHz and found the resonance response changing with lateral dot size. Whereas Raabe et al. [9] used small observation time after field pulse (1-14 ns), lower than the core oscillations period (estimated as ~ 30-40 ns for square thickness 30 nm and width 6 μm). This is the reason why they were unable to detect the core position oscillations. We measured initial core shift **X** during the field pulse ($t_1$ <0, see Fig. 2(A)-(B)) to be perpendicular to the pulse field direction.

In summary, we have experimentally detected the dynamics of magnetic vortices trapped in circular micron-size ferromagnetic dots using the XMCD/PEEM pump-probe technique. The eigenfrequency scales as a function of the dot geometrical aspect ratio. The measured frequencies are in agreement with analytical calculations that rely only on the known dot properties, such as the dot diameter and thickness, and the saturation magnetization. The observed vortex core position oscillations are interpreted as the translational mode of the magnetic vortex motion around the equilibrium position induced by a gyroforce and a dynamic magnetostatic restoring force.

The work was supported by the US Department of Energy, BES Materials Sciences under contract W-31-109-ENG-38.

**References**

[1] J. Martin, J. Nogues, K. Liu, J. Vicent, and I. Schuller, *J. Magn. Magn. Mat*. **256,** 449 (2003).

[2] R. P. Cowburn, D. K. Koltsov, A. O. Adeyeye, M. E. Welland, D. M. Tricker, *Phys. Rev. Lett.* **83,** 1042 (1999).

[3] K. Yu. Guslienko, V. Novosad, Y. Otani, H. Shima, K. Fukamichi, *Phys. Rev.* **B 65,** 024414 (2002).

[4] T. Shinjo, T. Okuno, R. Hassdorf, K. Shigeto, and T. Ono, *Science* **289,** 930 (2000).

[5] M. Schneider, H. Hoffman, and J.Zweck, *Appl. Phys. Lett.* **77**, 2909 (2000).

[6] A. Wachowiak et al., *Science* **298**, 577 (2002).

[7] S.-B. Choe et al., *Science* **304,** 420 (2004).

[8] A. Puzic et al., *J. Appl. Phys.* **97**, 10E704 (2005).

[9] J. Raabe et al., *Phys. Rev. Lett.* **94**, 217204 (2005).

[10] K.Yu. Guslienko et al., *Appl. Phys. Lett.* **78,** 3848 (2001).

[11] K. Yu. Guslienko et al., *J. Appl. Phys*. **91,** 8037 (2002).

[12] V. Novosad et al., *Phys. Rev*. **B 66**, 052407 (2002).

[13] L.Giovannini et al., *Phys. Rev*. **B 70,** 172404 (2004).

[14] B.A. Ivanov and C.E. Zaspel, *Phys. Rev. Lett*. **94,** 027205 (2005).

[15] K.Y. Guslienko, W. Scholz, R.W. Chantrell, and V. Novosad, *Phys. Rev.* **B 71**, 144407 (2005).

[16] X. Zhu, Z. Liu, V. Metlushko, P. Grütter, and M. R. Freeman, Phys. Rev. **B 71**, 180408 (2005).

[17] J. P. Park, P. Eames, D. M. Engebretson, J. Berezovsky, P. A. Crowell, *Phys. Rev*. **B 67**, 020403 (2003).

[18] V. Novosad et al., *Phys. Rev.* **B 72**, 024455 (2005).

[19] A. A. Thiele, *Phys. Rev. Lett.* **30,** 230 (1973).

[20] A.H. Lumpkin, F. Sakamoto, and B.X. Yang, The Proceedings of 2005 Particle Accelerator

Conference (PAC-05), submitted.

[21] N. Manton, and P. Sutcliffe, *Topological Solitons* (Cambridge University Press, Cambridge, 2004).

[22] A.M. Kosevich, B.A. Ivanov and A.S. Kovalev, *Phys. Reports* **194**, 117 (1990)

[23] W. Doering, J. Appl. Phys. **39**, 1006 (1968); A.A. Belavin and A.M. Polyakov, *JETP Lett*. **22**, 245 (1975).

[24] D.L. Huber, *Phys. Rev*. B **26**, 3758 (1982).

[25] G.M. Wysin, *Phys. Rev*. **B 54** 15156 (1996); ibid. **B 63**, 094402 (2001).

[26] K.Yu. Guslienko and A.N. Slavin, *Phys. Rev*. **B 72**, 014463 (2005).

**Figure captions**

Fig. 1. Schematic drawing of the time resolved PEEM setup. The figure on the lower right displays the waveform of the magnetic field generated by the current pulses.

Fig.2. (A) and (B), respectively, show the time dependence of the *Y* and *X* displacement of the vortex core of a 6.3 micron diameter FeNi cylindrical dot. At the top are PEEM images of the dot as a function of time. The corresponding data in (A) are related to the images by the letters (a)-(l). The arrows in image (a) indicate the direction of the magnetization.

Fig. 3. Time dependence of the *Y* displacement of the vortex core of 5.3 and 4.3 micron diameter FeNi cylindrical dots.

Fig. 4. Comparison of the XMCD/PEEM experimental data (symbols) and analytical theory for the dependence of the eigenfrequency of the vortex translational mode on the dot aspect ratio $\beta=L/R$ of the cylindrical FeNi dots. The experimental frequencies are measured for the pulse widths of 19 (upper symbols) and 75 ns. The solid line is plotted according to Eq. (3). The dot thickness is *L*=30 nm and the diameters are 2R=4.3, 5.3 and 6.3 µm.

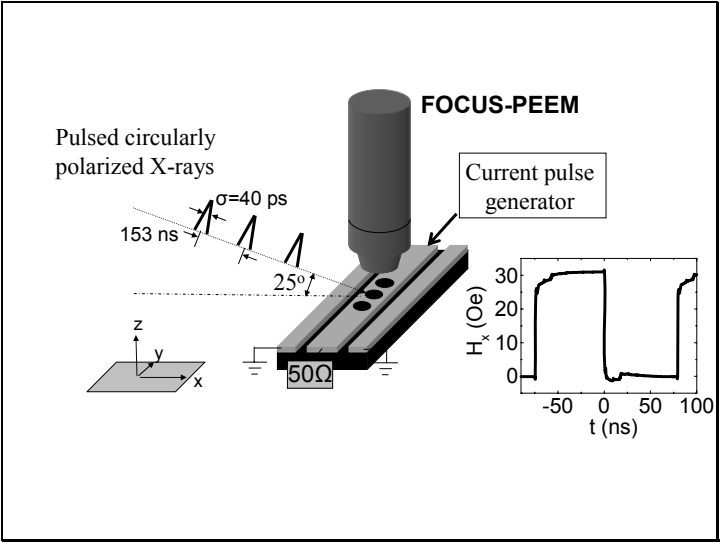

Fig. 1.

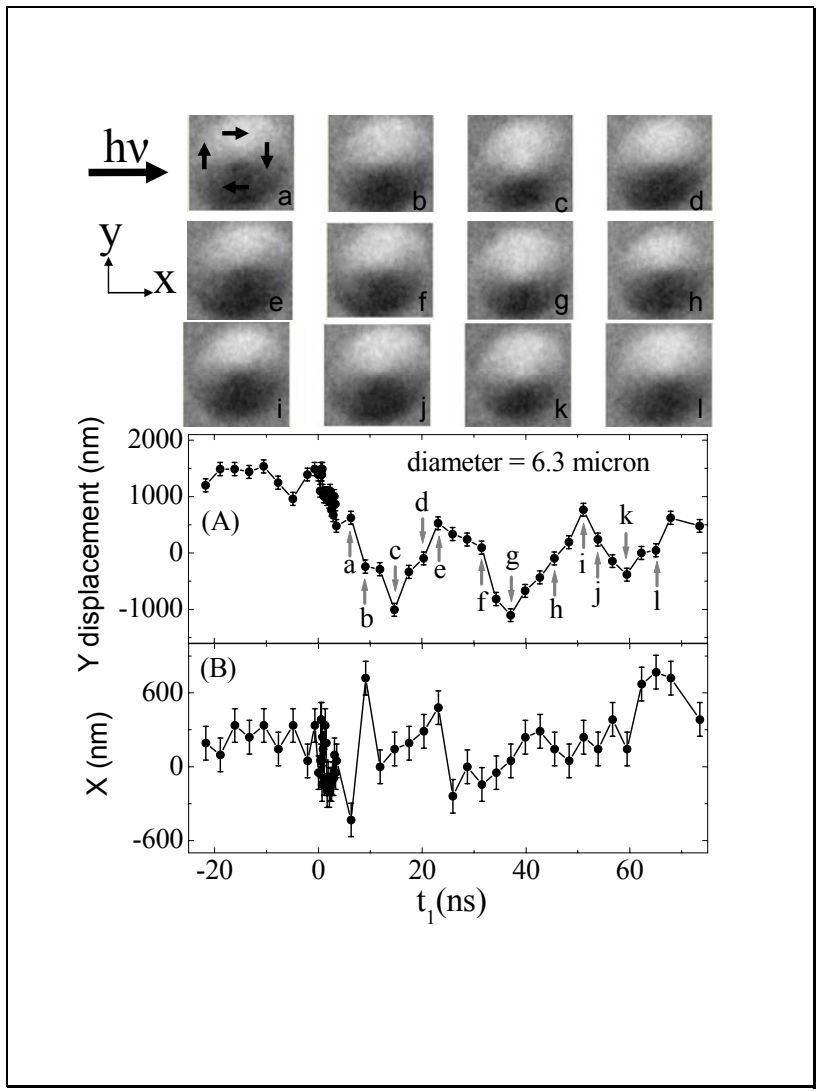

Fig. 2.

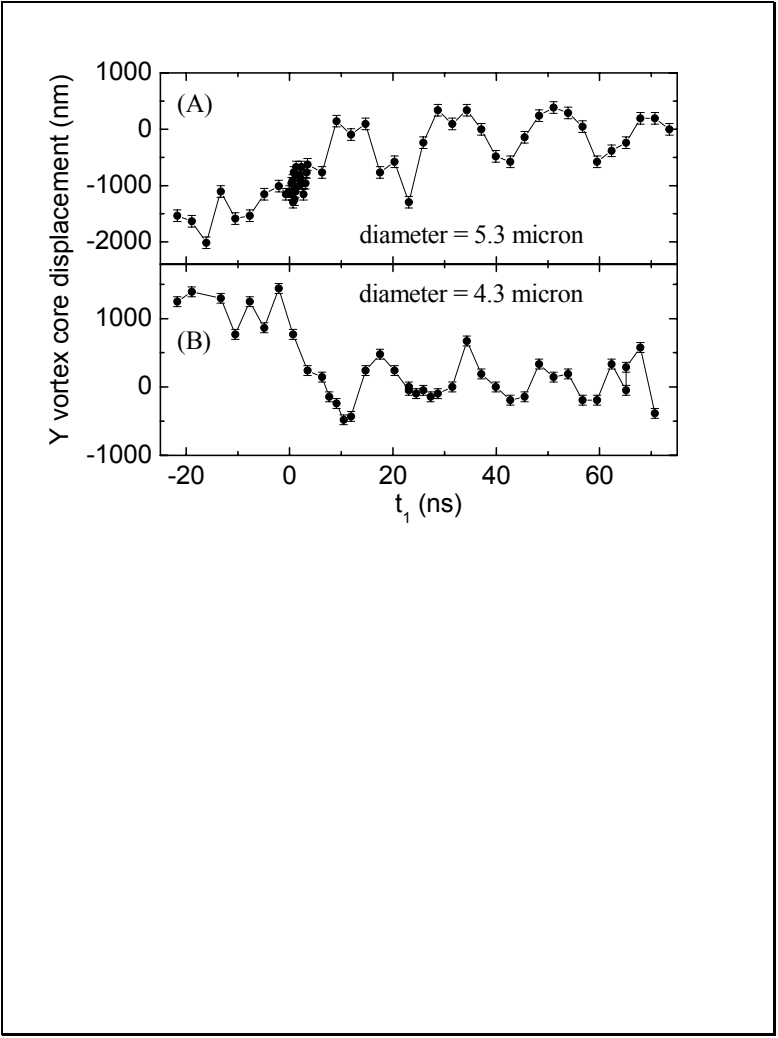

Fig. 3.

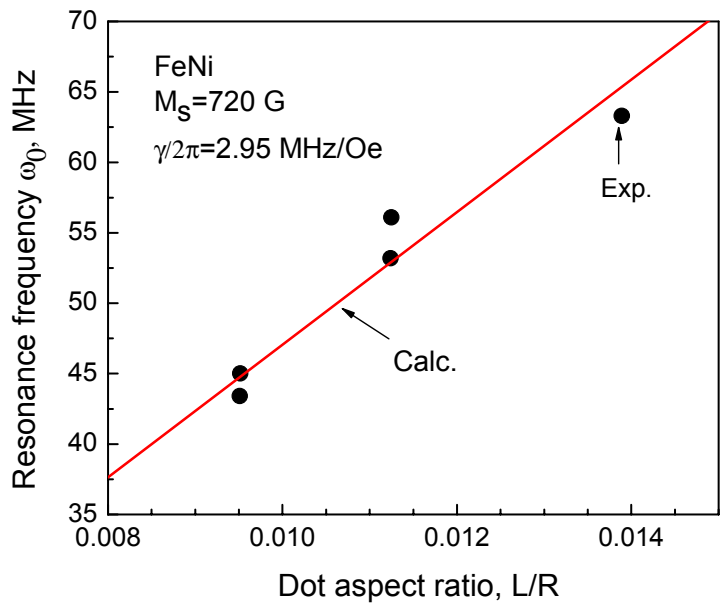

Fig. 4.